\documentstyle[12pt,epsfig]{article}
\def\lsim{\raise0.3ex\hbox{$<$\kern-0.75em\raise-1.1ex\hbox{$\sim$}}}
\def\gsim{\raise0.3ex\hbox{$>$\kern-0.75em\raise-1.1ex\hbox{$\sim$}}}
\begin {document}

\noindent DFTT 24/2001\\
FISIST/12-2001/CENTRA\\
28 August 2001

\vskip 2.5 truecm

\begin{center}
{\Large {\bf INCLUSIVE DENSITY OF SECONDARIES IN HEAVY ION COLLISIONS}} \\
\vskip 1.5 truecm
{\bf J.Dias de Deus and Yu. M. Shabelski$^*$}\\
\vskip 0.5 truecm
CENTRA, Instituto Superior T\'ecnico, 1049-001 Lisboa, Portugal 
\vskip 0.75 truecm
{\bf R. Ugoccioni}\\
\vskip 0.5 truecm
Dipartimento di Fisica Teorica and INFN -- Sezione di Torino,
10134 Torino, Italy
\end{center}
\vskip 1.5 truecm
\begin{center}
{\bf ABSTRACT}
\end{center}

The inclusive density of charged secondaries produced in high energy heavy 
ion collisions are calculated in the framework of multiple scattering 
theory and quark-gluon string model. The obtained results are in 
agreement with the data only assuming rather large inelastic 
screening/percolation effects. We predict the total saturation of relative 
inclusive densities in dependence on the number of interacting nucleon.
Predictions for LHC are also given.   
   
\vskip 1.5 truecm

\noindent $^*$Permanent address: Petersburg Nuclear Physics Institute,
Gatchina, St.Petersburg, Russia \\
E-mail: shabelsk@thd.pnpi.spb.ru

\newpage
\pagestyle{plain}
\noindent{\bf 1. Introduction}
\vskip 0.5 truecm

The processes of high energy heavy ion collisions can be considered in the 
framework of multiple scattering theory. Many references can be found in 
reviews \cite{PR,BSh,AP,JSh}. Contrary to the case of hadron-nucleus 
interactions, in the case of heavy ions it is impossible to sum 
analytically\footnote{It can be done with the help of Monte Carlo 
calculation of multidimensional integral \cite{KSh}.} the contributions of 
all diagrams of Glauber approach. 

However, the contributions of the most important diagrams can be accounted 
for analytically, that allows one to calculate the integral cross sections
of different processes, distributions on the number of interacting 
nucleons, multiplicities of secondaries and their inclusive distributions
in reasonable agreement with the data. 

It is well-known that in high energy hadron-nucleus collision there exists
inelastic screening \cite{Gr,KM} which is confirmed experimentally, 
especially for the case of hadron-deuteron interactions. The same 
inelastic screening should exist in high energy heavy ion collision. This 
effect is very small for integrated cross sections (because many of them 
are determined by geometry), but it is very important \cite{CKT} for the 
calculations of secondary multiplicities and inclusive densities.
Similar results can be obtained \cite{JU,APS} in the framework of string 
fusion \cite{ABP}, or percolation \cite{BP} models where string 
fusion/percolation effects directly correspond \cite{APSh} to the pomeron 
interactions. 

In the present paper we will give the theoretical estimations for 
relative inclusive densities, integrated multiplicities and rapidity 
distributions using both Gribov's reggeon diagram technique \cite{Gr1} and 
percolation approach as well as Quark-Gluon String Model (QGSM)
\cite{KTM,KTMSh}.

We will also compare the above estimates with results from the Dual String 
Model (DSM) \cite{JU,JU:2} and find them remarkably similar, although the 
DSM is based on a different approach: it takes as input quantities 
measured in $p\bar p$ and/or $pp$ collisions, and screening is realised 
via fusion of strings exchanged in partonic sea collisions, which can lead 
to percolation effects \cite{JUR:Jpsi}. 

We will show that there exists a qualitative difference between 
secondary production at CERN SpS and RHIC energies and this difference can 
be described only by accounting for rather large inelastic 
screening/percolation/string fusion effects. These effects decrease the 
inclusive density about two times, that is in agreement with previous 
estimations \cite{CKT}. 

For the distributions calculated in the present paper in DSM it is string 
fusion that leads to saturation with the number of participant nucleons, 
percolation phase transition is not felt.

\noindent{\bf 2.Spectra of secondaries and multiplicities in heavy ion 
collisions}

We start with standard multiple scattering theory consideration without 
inelastic screening. At high energies the spectrum of a secondary particle 
$h$ produced in the central rapidity region of nucleus $A$ - nucleus $B$ 
collision is proportional to the same spectrum in $NN$ collision 
\cite{Sh1}
\begin{equation}
\frac{d\sigma (AB \to hX)}{dy} = A\cdot B \frac{d\sigma (NN \to hX)}{dy}
\end{equation}

However at existing energies the right-hand side of this expression is 
significantly smaller due to energy conservation corrections. The 
values of these corrections depend on the classes of multiple scattering 
diagrams which are taken into account. This problem was analysed in 
\cite{Sh2} and the results obtained in the so-called rigid target 
approximation \cite{Alkh,GCSh} were in agreement \cite{GCSh} with the low 
energy ($\sim$ 4 GeV per nucleon) experimental data on heavy ion 
collisions. 
 
Rigid target approximation corresponds to the account for the 
contributions of diagrams shown in Fig. 1. Every nucleon of a beam nucleus 
\begin{figure}
\centerline{\epsfig{file=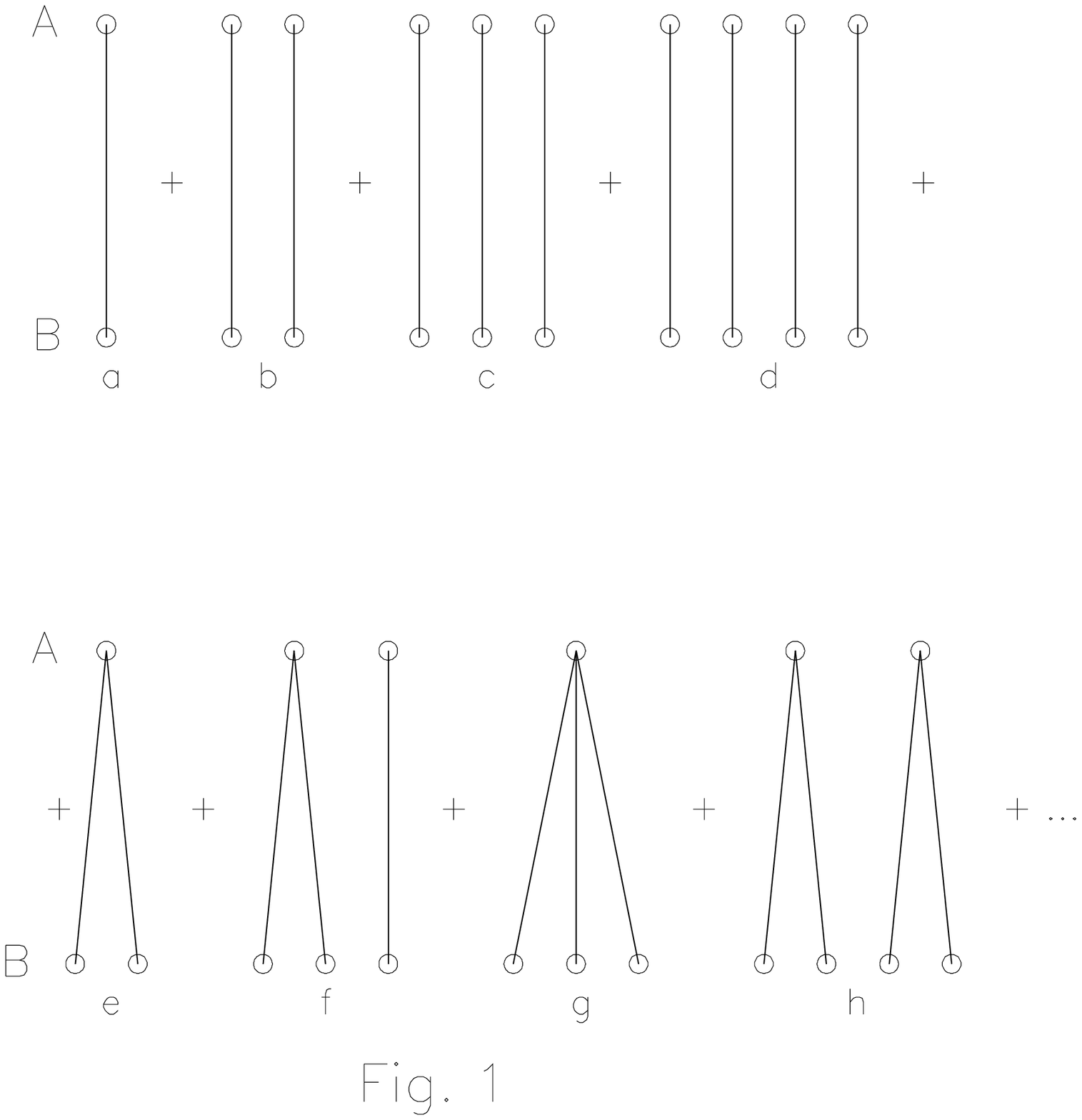,width=4in}}
\caption{Diagrams corresponding to the rigid target approximation for  
nucleus $A$ -- nucleus $B$ interaction. Only interacting nucleons are
shown as the open points.}
\end{figure}
$A$ can interact several times with a target nucleus $B$, however every 
nucleon of $B$ nucleus can interact only one time. This asymmetrical 
picture seems to be reasonable for nucleus $A$ fragmentation, because the 
main part of shadow corrections is accounted for. 
For fragmentation of nucleus 
$B$ the result seems to be inconsistent. So the inclusive spectrum of 
secondary $h$ from $AB$ collision can be written \cite{Sh2} as  
\begin{eqnarray}
\frac1{\sigma^{prod}_{AB}} \frac{d\sigma (AB \to hX)}{dy} & = &
\theta (y) R^h_A(y) <\!N_A\!> \frac1{\sigma^{prod}_{NB}} 
\frac{d\sigma (NB \to hX)}{dy} + \nonumber \\
& + & \theta (-y) R^h_B(-y) <\!N_B\!> \frac1{\sigma^{prod}_{NA}} 
\frac{d\sigma (NA \to hX)}{dy} \;,
\end{eqnarray}
where $y$ is the rapidity of secondary $h$ in c.m.s., $<\!N_A\!>$ and 
$<\!N_B\!>$ are the averaged numbers of participating nucleons for every 
incident nuclei and the functions $R^h_{A,B}(y)$ account for the energy 
conservation effects. These last quantities can be found from the 
calculated inclusive spectra in $NA$ and $NB$ collisions, and they differ 
from unity about 20\% at CERN $SpS$ energy and about 10\% at RHIC energy.

Now we have to calculate the inclusive cross sections of secondary 
particle $h$ in $NA$ and $NB$ collisions. It can be done with the help of 
the quark--gluon string model (QGSM) \cite{KTM,KTMSh}. This model is
based on the Dual Topological Unitarization (DTU) and it describes quite
reasonably many features of high energy production processes including 
the inclusive spectra of different secondary hadrons, their 
multiplicities, KNO--distributions, etc., both in hadron--nucleon and 
hadron--nucleus collisions \cite{KTM,KTMSh,Sh3}. 

High energy interactions are considered in QGSM as proceeding via the 
exchange of one or several pomerons and all elastic and inelastic 
processes result from cutting through or between pomerons \cite{AGK}. The 
possibility of exchanging a different number of pomerons introduces 
absorptive corrections to the cross sections which are in agreement with 
the experimental data on production of hadrons consisting of light quarks. 

Each pomeron corresponds to a cylindrical diagram, Fig. 2a, and thus, 
when cutting a pomeron two showers of secondaries are produced (Fig. 2b). 
The inclusive spectra of secondaries are determined by the convolution of 
diquark, valence and sea quark distributions $u(x,n)$ in the incident 
particles and the fragmentation functions $G(z)$ of quarks and diquarks 
into secondary hadrons. Both the initial quark distributions and the 
fragmentation functions of quarks and diquarks are constructed using the 
reggeon counting rules \cite{Kai}.

\begin{figure}
\centerline{\epsfig{file=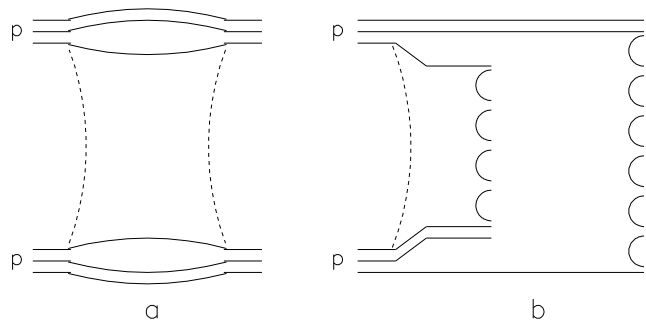,width=4in}} 
\caption{Cylindrical diagram which corresponds to the one-pomeron 
exchange contribution to elastic $pp$ scattering (a) and its cut which 
determines the contribution to inelastic $pp$ cross section (b).}
\end{figure}

The diquark and quark distribution functions depend on the number $n$ of cut
pomerons in the considered diagram. In the following we use the formalism of
QGSM. In the case of a nucleon target the inclusive spectrum of a secondary
hadron $h$ (its rapidity, $y$, or Feynman-$x$ distribution) has the form 
\cite{KTM}: 
\begin{equation}
\frac{dn}{dy} = 
\frac1{\sigma_{inel}} \frac{d\sigma}{dy} =
\frac{x_E}{\sigma_{inel}} \frac{d\sigma}{dx_F} 
=\sum_{n=1}^{\infty}w_{n}\phi_{n}^{h}(x) \; , 
\end{equation}
where the functions $\phi_{n}^{h}(x)$ determine the contribution of
diagrams with $n$ cut pomerons and $w_{n}$ is the probability of this
process. Here we neglect the contributions of diffraction dissociation
processes which are comparatively small in most of the processes
considered below. It can be accounted for separately \cite{KTM}.

For $pp$ collisions
\begin{equation}
\phi_{pp}^{h}(x) = f_{qq}^{h}(x_{+},n)f_{q}^{h}(x_{-},n) +
f_{q}^{h}(x_{+},n)f_{qq}^{h}(x_{-},n) +
2(n-1)f_{s}^{h}(x_{+},n)f_{s}^{h}(x_{-},n)\ \  ,
\end{equation}
\begin{equation}
x_{\pm} = \frac{1}{2}[\sqrt{4m_{T}^{2}/s+x^{2}}\pm{x}]\ \ ,
\end{equation}
where $f_{qq}$, $f_{q}$ and $f_{s}$ correspond to the contributions of
diquarks, valence and sea quarks respectively. They are determined by
the convolution of the diquark and quark distributions with the
fragmentation functions, e.g.,
\begin{equation}
f_{q}^{h}(x_{+},n) = \int_{x_{+}}^{1} u_{q}(x_{1},n)G_{q}^{h}(x_{+}/x_{1})
dx_{1}\ \ .
\end{equation}

The diquark and quark distributions as well as the fragmentation functions 
are determined from Regge intercepts. Their expressions are given in 
Appendix.

The calculation of the inclusive spectra on nuclear targets is similar to
Eq. (3). The only difference is that in the case of interaction, say, with
three target nucleons and with exchange by five pomerons, it is necessary 
to account for all possible permutations. 

The multiplicities of the produced secondaries can be obtained by 
integrating the inclusive spectra.

\vskip 0.9 truecm
\noindent{\bf 3. Comparison with data}
\vskip 0.5 truecm

Many examples of the QGSM description of experimental data can be found in 
\cite{KTM,KTMSh,Sh3,ACKSh,KP}. Now we are more interested in particle 
yields in central (mid-rapidity) region. The data on secondary $\pi^{\pm}$ 
and $K^{\pm}$ production in $pp$ collisions at ISR energies \cite{ISR} at 
$90^o$ in c.m.s. are presented in Figs. 3a and 3b. We slightly change the 
pomeron theory parameters, namely its intercept was assumed to be
$\alpha_P(0) = 1 + \Delta$ with $\Delta = 0.09$. This values of $\Delta$ 
results in slightly better description of the ISR data. It is practically 
the same as Donnachie--Landshoff \cite{DL} value $\Delta = 0.08$ and 
rather smaller than the value $\Delta = 0.17$ which was obtained in 
\cite{Likh}. The description of secondary protons and antiprotons can be 
found in \cite{ACKSh}. One can see that these data are described by QGSM 
quite reasonably.

\begin{figure}
\begin{minipage}[b]{7in}
\centerline{\epsfig{file=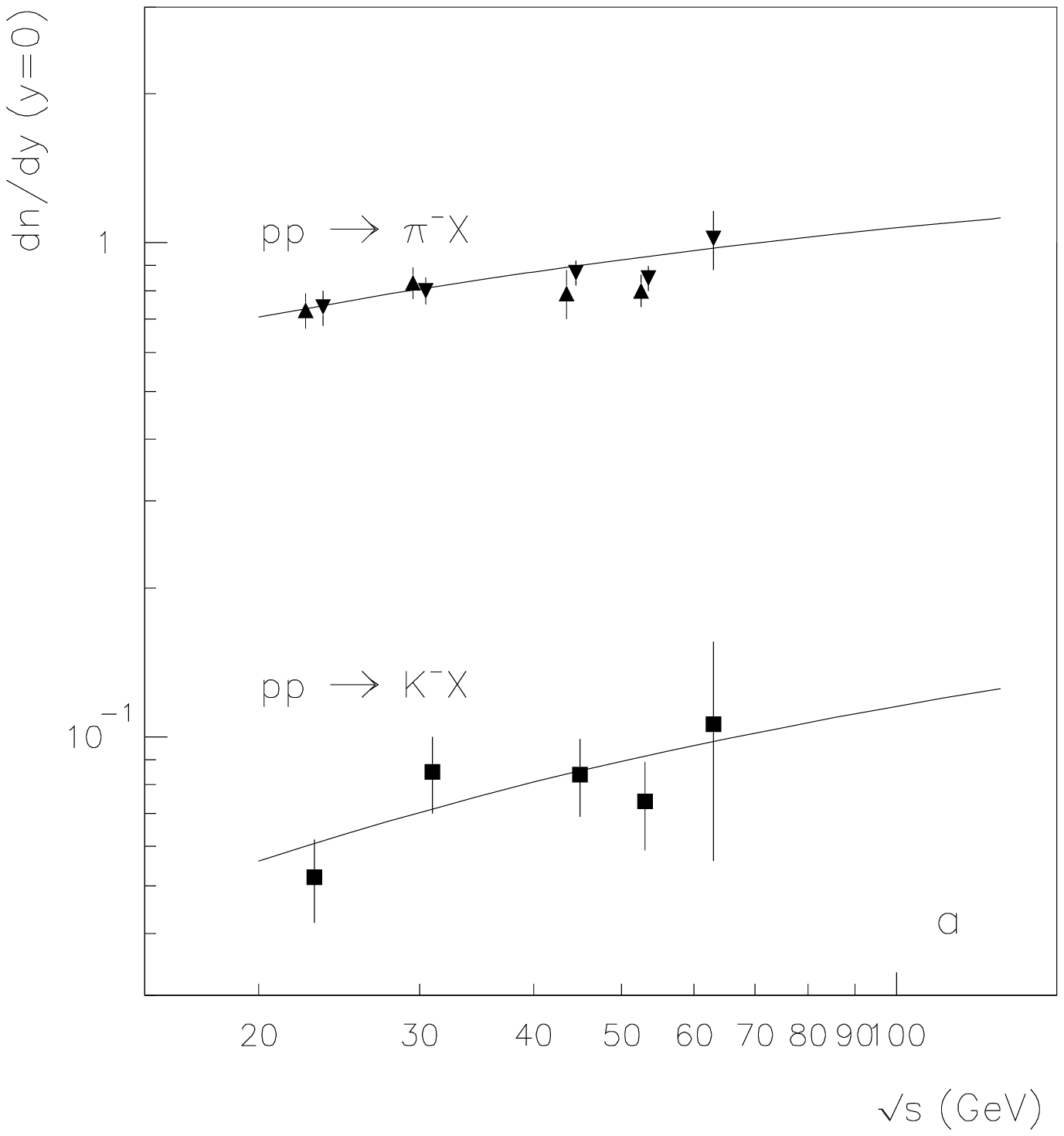,width=3in}   
\hspace{0.2cm}\epsfig{file=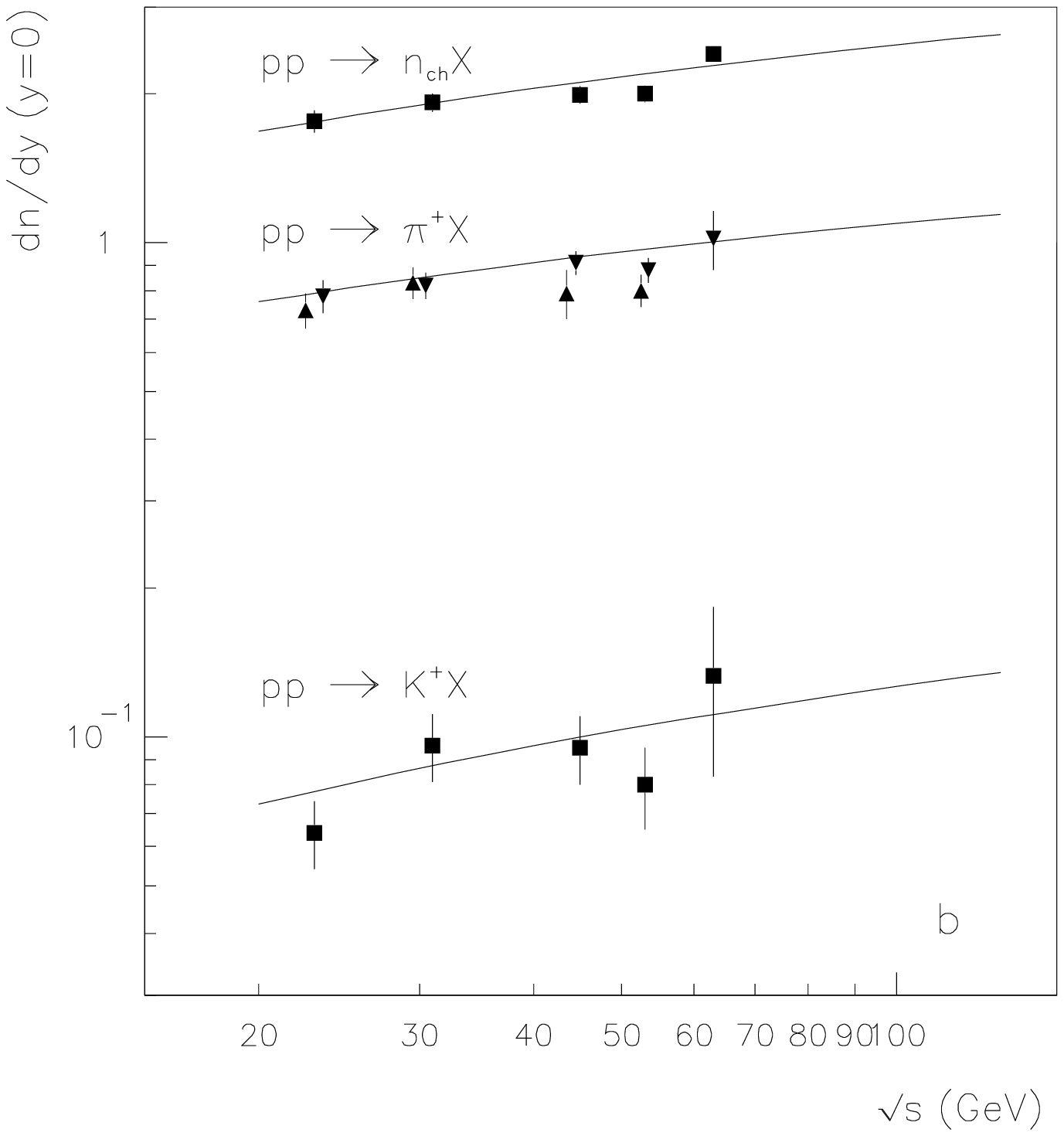,width=3in}}
\hfill
\end{minipage}
\caption{Secondary negative (a) and positive (b) pion and kaon yields at 
ISR energies \cite{ISR} at $90^o$ in c.m.s. together with the data for all 
charged secondaries and with QGSM predictions.}
\end{figure}

Now we can consider the secondary production in heavy ion collisions.
The results of calculations of inclusive density of charged secondaries in 
mid-rapidity region for CERN SpS energy are in agreement with the data, 
whereas for RHIC energies are more than two times larger than the RHIC 
experimental data, and it is impossible to change QGSM parameters to have 
significantly better description for both $pp$ and heavy ion cases. 

This disagreement is rather expected. At comparatively low energies (not 
higher than several GeV per nucleon) the heavy ion collisions can be 
considered with the help of standard Glauber approximation, and the 
inclusive spectrum of any secondary $h$ produced in the central 
(mid-rapidity) region is described by the diagram shown in Fig. 4a, i.e. 
by the contribution of a single nucleon-nucleon blob. This diagram 
immediately gives Eq. (1) for heavy ion inclusive cross section. The 
contributions of all other diagrams cancel each other due to AGK cutting 
rules \cite{AGK}.

\begin{figure}
\centerline{\epsfig{file=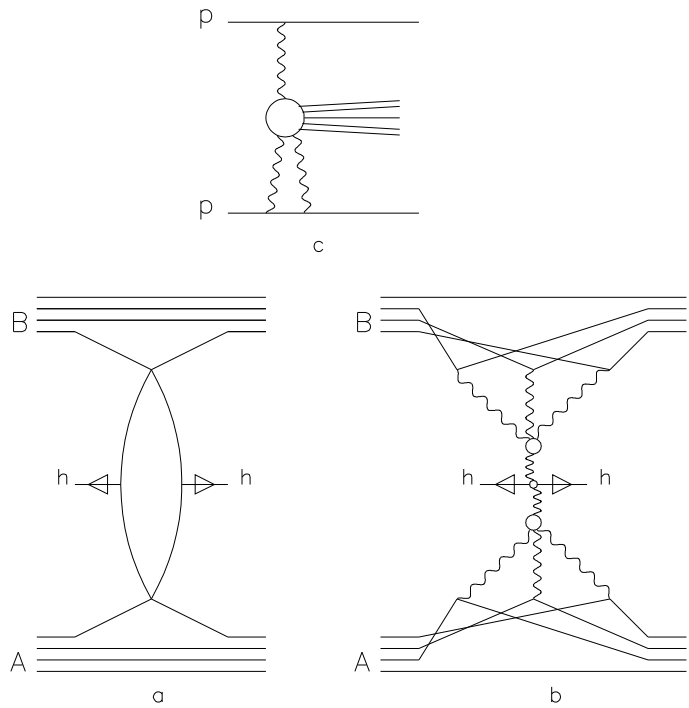,width=4in}}
\caption{Diagrams for inclusive cross sections for nucleus $A$--nucleus 
$B$ collisions in Glauber approximation (a) and with accounting for the 
interactions of pomerons (shown by wave curves) (b). An example of 
inelastic processes of $pp$ interactions which determine one of the 
needed vertices of the multipomeron interactions (c).} 
\end{figure}

At high energies the new diagrams appear which include the interactions of 
pomerons. The simplest example of such diagrams is so-called 
triple-pomeron diagram which describe the diffractive production of a 
large mass $M$ jet in $pp$ interactions. At low energies the contribution 
of such diagrams to hadron-nucleus or nucleus-nucleus interaction is 
suppressed by the longitudinal part of nuclear form factor \cite{CKT} 
\begin{equation}
F_A(t_{min}) \approx \exp{(R^2_A t_{min}/3)} 
\end{equation}
for Gaussian distribution of nuclear density, where 
\begin{equation}
t_{min} \approx (m_N M^2/s)^2 \;. 
\end{equation}
When energy becomes high enough, the value of $t_{min}$ becomes very 
small. So the discussed contribution can be significant and namely these 
diagrams determine the inelastic screening corrections to hadron-nucleus 
cross sections. 

One example of a diagram with pomeron interaction for heavy ion 
interaction is shown in Fig. 4b. Contrary to hadron-nucleus case the
contribution of such diagram can be estimated from the processes of high
mass jet production in mid-rapidity region and with two large rapidity 
gaps, see for example Fig. 4c. The numerical contribution of all such 
diagrams is rather unclear because the number of diagrams is very large 
and the vertices of multipomeron interactions are unknown. For example, in 
\cite{CKT} the Shwimmer model was used for the numerical estimations. 
However all such estimation are model dependent.  

The numerically significant contribution of the considered diagrams to 
inclusive density is suppressed quadratically, by both longitudinal form 
factors, $F_A(t_{min})$ and $F_B(t_{min})$. So we can observe their 
influence at energies quadratically higher in comparison with the energies 
region where the inelastic screening effects are observed in 
hadron-nucleus scattering. One can see that the RHIC energies are of the 
needed order of magnitude.

The reasonable possibility to estimate the contribution of the diagrams 
with pomeron interaction comes from percolation theory. We assume that if 
two or several pomerons are overlapping, they become one pomeron. It means 
that the inclusive density is saturated, it reaches its maximal value at 
given impact parameter. This approach has one free parameter - the 
critical number of pomerons in one squared fermi. Technically it is more 
simple to bound the maximal number of pomerons, $n_{max}$, which can be 
emitted by one participating nucleon for the given pair of colliding 
nuclei. All model calculations become rather simple because above the 
critical value every additional pomeron cannot contribute to the 
inclusive spectrum. 

\begin{figure}
\centerline{\epsfig{file=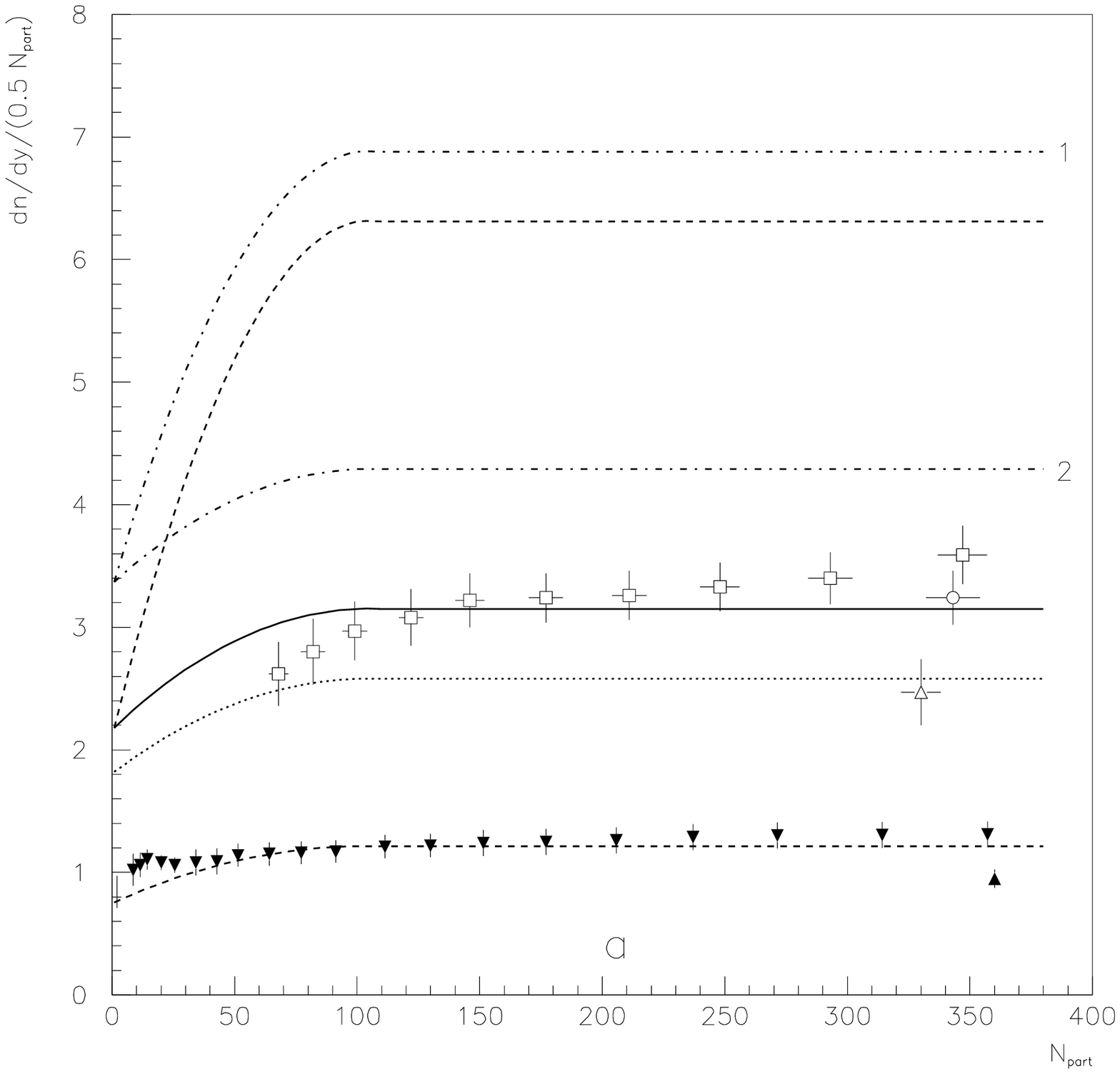,width=3in}
\hspace{0.2cm}\epsfig{file=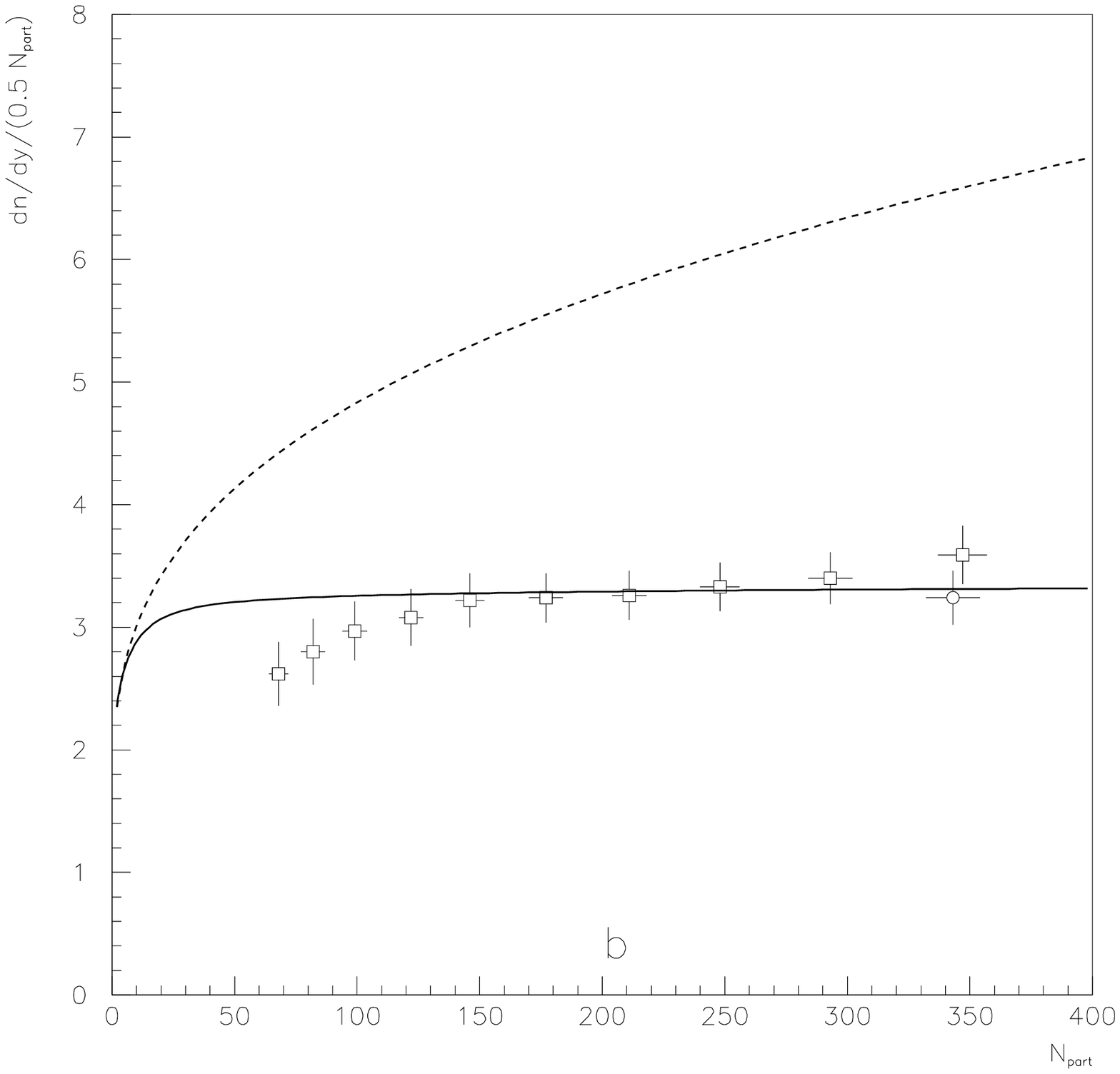,width=3in}}
\caption{Relative inclusive densities of secondaries for $Pb-Pb$ 
collisions at $\sqrt{s} = 17.3$ GeV per nucleon (black points, 
\cite{SpS,SpS1,Phob}, multiplied by 1/2) and for $Au-Au$ at $\sqrt{s} = 
130$ GeV per nucleon (open points, \cite{Phob,Phen}) and  at $\sqrt{s} = 
56$ GeV per nucleon (open triangle, \cite{Phob}). In Fig. 5a the solid curve 
shows the results of calculations accounting for the interactions of 
pomerons with $n_{max}=1.67$. Dashed curves present the results of the 
Glauber approximation (without percolation effects) for $\sqrt{s} = 17.3$ 
GeV per nucleon (lower curve) and $\sqrt{s} = 130$ GeV per nucleon (upper 
curve). Dotted curve shows our predictions for $Au-Au$ at 56 GeV per 
nucleon with $n_{max}=1.67$, and dash-dotted curve 1 and 2 for $Pb-Pb$ 
collisions at 5.5 TeV per nucleon with $n_{max}=1.67$ and $n_{max}=1$, 
respectively. In Fig. 5b the results of DSM calculations at $\sqrt{s} = 
130$ GeV per nucleon with and without string fusion are shown by solid and 
dashed curves.}
\end{figure}

The Dual String Model on the other hand is based essentially on Dual 
Parton Model ideas \cite{DPM} with the inclusion of strings \cite{ABP} 
which may interact in the transverse plane and fuse, thus reducing their 
contribution to the final state. Hadrons are considered as made up of 
constituent quarks (valence and sea quarks): the valence-valence diagram 
corresponds to single inelastic scattering and the wounded nucleon model 
\cite{WNM}, while the sea-sea diagram (including gluons) corresponds to 
additional inelastic multiple scattering contributions. These 
contributions may be internal, parton multiple scattering within the 
original valence-valence contribution, or external, involving other
nucleons. It should be noticed that the DSM applies only to symmetrical
collisions (A=B).

The main equation of the DSM was originally written for the
rapidity particle density as
\begin{equation}
  \left.\frac{dN}{dy}\right|_{N_AN_A} = 
                N_A \left[ 2 + 2(k-1)\alpha\right] h
            + (\nu - N_A)  2k\alpha h ,
	  \label{eq:Armesto}
\end{equation}
where $h$ is the height of the rapidity plateau for 
valence-valence collisions, $\alpha h$ the
height of the sea-sea plateau, $2k$ the average number of strings
produced in a nucleon-nucleon collision and $N_A$ is the number of
participating nucleons from nucleus $A$; finally
$\nu$ is the average number of nucleon-nucleon collisions, which from
elementary multiple scattering arguments \cite{Armesto:SFM} 
satisfies at high energy $\nu \sim N_A^{4/3}$.
Notice that the first term on the right-hand side is exactly
$N_A$ multiplied by the contribution of one nucleon-nucleon collision,
while the second term results from sea-sea type collisions.
One should notice that the number of nucleon-nucleon collisions is
$N_A+(\nu-N_A) = \nu$ and the number of strings is $N_A\left[ 2
+2(k-1)\right] + (\nu - N_A) 2k = 2k\nu$.
Assuming that $h$ and $\alpha$ are energy independent (constant
plateaus) the energy dependence of ${dN}/{dy}|_{pp}$, taken
in this model by interpolating experimental data, fixes the energy
dependence of $k$, as explained in ref.~\cite{JU}, where
it was found that $\alpha = 0.05$ and $h=0.75$; all results of the
DSM shown in this paper are obtained with such values.

By integrating Eq.~(\ref{eq:Armesto}) over the whole rapidity interval
predictions are obtained from the DSM for the
average multiplicity: the model's result is clearly linear
but it overestimates the data. As argued in \cite{JU:2},
the assumption that the sea-sea plateau is proportional to
the valence-valence plateau is not valid in the fragmentation region,
effectively one can say that $\alpha=0$ there.
Fitting the experimental data allows us to
estimate that Eq.~(\ref{eq:Armesto}) is valid over the central region
corresponding to 70\% of total phase space.

The results of calculations of the inclusive densities of the produced 
charged secondaries per one pair on interacting nucleons 
$dn_{ch}/dy/(0.5 N_{part})$, where $N_{part} = N_A$ with the help of 
Eq. (2) and QGSM are shown in Fig. 5a. We present the experimental data 
for $Pb-Pb$ collisions at $\sqrt{s} = 17.3$ GeV per nucleon 
\cite{SpS,SpS1,Phob} for $\vert y \vert < 1$ and for $Au-Au$ at $\sqrt{s} 
= 130$ GeV per nucleon \cite{Phob,Phen} and at $\sqrt{s} = 56$ GeV per 
nucleon for $\vert \eta \vert < 1$\footnote{We accounted for the 
difference between rapidity and pseudorapidity distributions.}. One can 
see that CERN SpS data are described reasonably without any additional 
screening (percolation). It was shown also in \cite{JSh} and it seems to 
be natural because the suppression effects of $t_{min}$ discussed above 
should be rather large at comparatively small energy of CERN SpS.

However, the same calculation at RHIC energy $\sqrt{s} = 130$ GeV per 
nucleon gives the relative inclusive density two times larger (upper 
dashed curve in Fig. 5a) than the data, that is in numerical agreement 
with \cite{CKT}. The agreement with the data can be obtained only with 
suppression of multipomeron contributions, namely by using $n_{max} = 
1.67$ for every interacting nucleons. The results of DSM calculations with 
and without string fusion are shown in Fig. 5b, and they are in agreement 
with the curves in Fig. 5a.  

The same value of $n_{max} = 1.67$ allows us to describe the PHOBOS point 
at 56 GeV (dotted curve in Fig. 5a). In the case of LHC energy $\sqrt{s} = 
5.5$ TeV per nucleon for $Pb-Pb$ collisions with $n_{max} = 1.67$ 
percolation effect decreases the relative inclusive density about 3 times 
(again similar to \cite{CKT} prediction). The recent result of PHOBOS 
Coll. \cite{Phob2} for central $Au-Au$ collisions at $\sqrt{s} = 200$ GeV 
per nucleon gives $d n_{ch}/d \eta = 650 \pm 35$ for $\vert \eta \vert < 
1$ that is in agreement with our result $d n_{ch}/d \eta \approx 600$.

However, the percolation effect at LHC energy can be even larger. The 
transverse range of a pomeron increases with incident energy as  
\begin{equation}
r^2_P \sim \alpha ' \ln{s/s_0} \;,
\end{equation}
that is well-known experimentally from the shrinkage of slope parameter in 
elastic hadron-nucleon scattering. So in the percolation picture the value 
of $n_{max}$ at higher energies should decrease. The predictions for the 
relative inclusive density at LHC energy with $n_{max} = 1$ (maximal 
percolation) are shown in Fig. 5a by dash-dotted curve 2. Now the 
percolation effect decreases the relative inclusive density about 5 times.

Let us note that Eq. (2) predicts independence of $dn_{ch}/dy/0.5 N_A$ on
the $N_A$ value for $N_A$ larger than $<\!N_A\!>$, that is confirmed by the 
data\footnote{In the presented approach we can calculate inclusive 
density for $N_A$ smaller than $<\!N_A\!>$ only in DSM.}. Really some 
rather weak dependence can exist because in central collisions every 
participating nucleon interacts more "centrally" on the average in 
comparison with minimum bias events. However this numerically small effect 
($\sim$10-20\% \cite{PSh}) should be practically suppressed by percolation 
effects except of rather peripheral collisions, where the value of 
$N_{part}$ is small and percolation effects are not so important. 

The same calculations in the framework of DSM gives similar results as it 
is shown in Fig. 5b for PHENIX data at $\sqrt{s} = 130$ GeV. Notice that 
contrary to the case of QGSM approach with Eq. (2) there is no saturation 
with the number of participants without string fusion. 

In Fig. 6 we present the calculated values of the relative inclusive 
density for secondary production in $pp$ and $Pb-Pb$ collisions as the 
functions of initial energy. One can see that due to percolation of 
pomerons emitted by different nucleons with similar impact parameters,
the  relative inclusive density in $Pb-Pb$ collisions can be smaller than
in the $pp$ case. In the case of DSM model the growth with energy is more 
pronounced than in the QGSM due to the extrapolation of $pp$ data with the 
second power of $\log s$.

\begin{figure}
\centerline{\epsfig{file=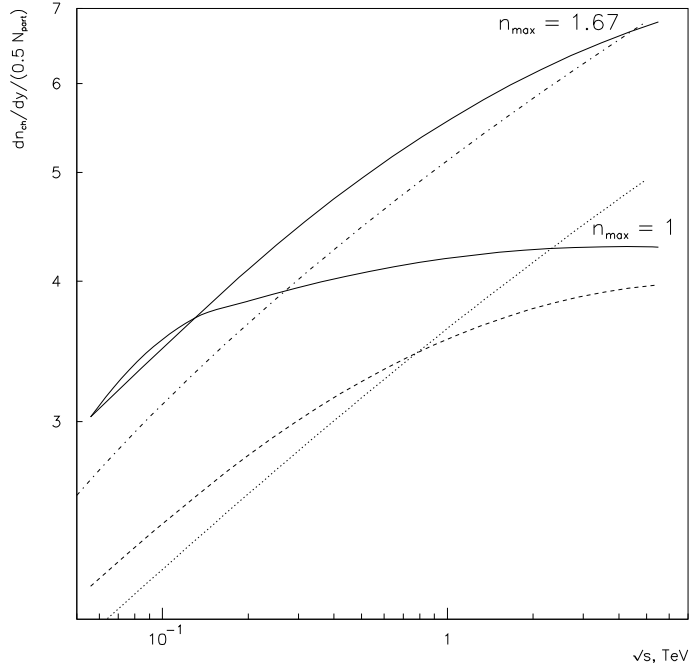,width=4in}}
\caption{Predictions for the relative inclusive charged particles density 
for secondaries within range $\vert y \vert < 1$ in $pp$ (dashed curve, 
QGSM and dotted curve, DSM) and $Pb-Pb$ (solid curve, QGSM and dash-dotted 
curves, DSM) collisions as the functions of initial energy. Two variants 
of solid curve correspond to percolation effects with $n_{max} = 1.67$ 
independently on energy and to its decrease from $n_{max} = 1.67$ to 
$n_{max} = 1$.}
\end{figure}

Now let us consider the data \cite{Phob1} on the total number of charged 
particles detected within range $-5.4 < \eta < 5.4$, i.e. in the region 
which exclude only high fragmentation and diffraction regions. 

\begin{figure}
\centerline{\epsfig{file=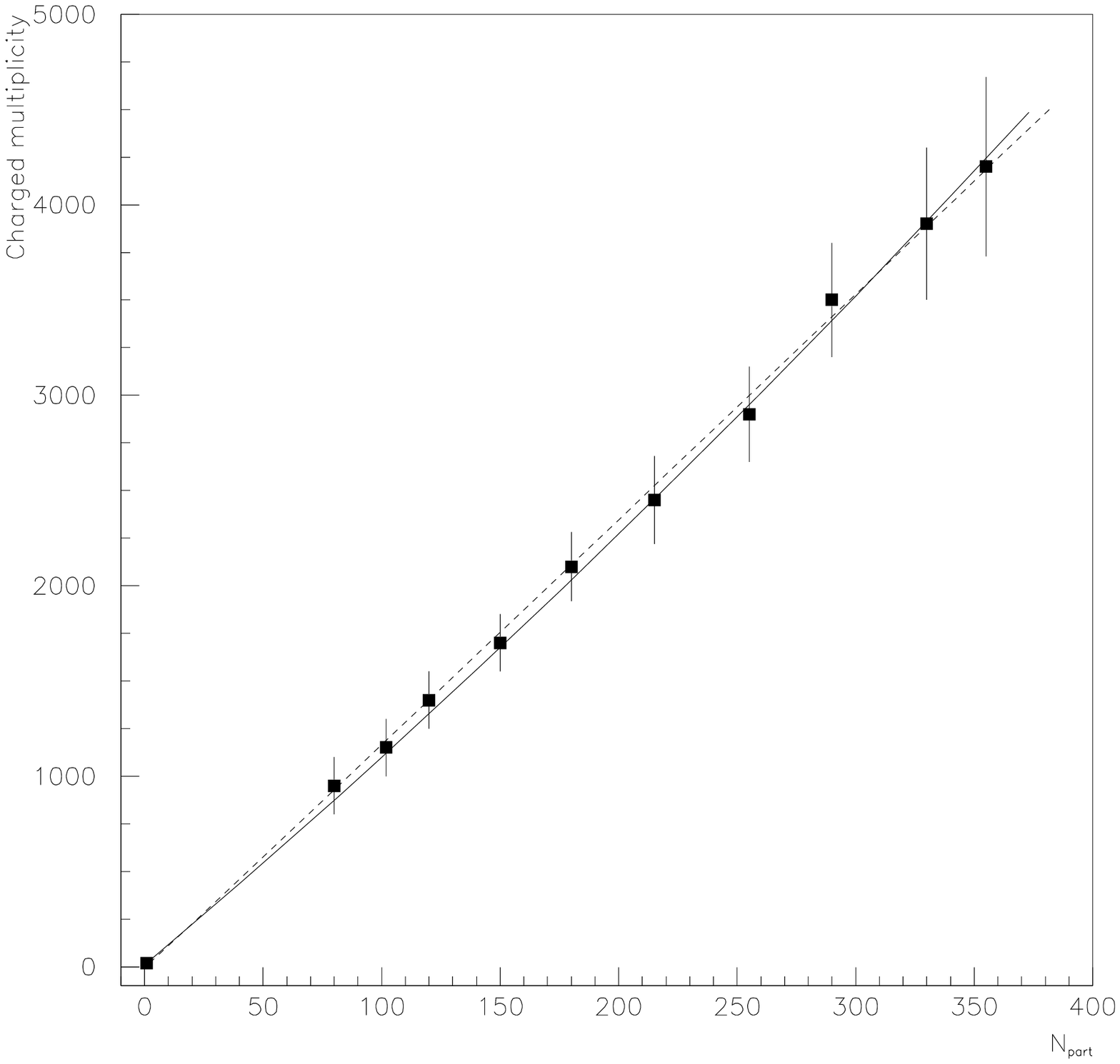,width=5in}}
\caption{Total number of charged particles detected within range $-5.4 < 
\eta < 5.4$ in $Au-Au$ collisions at $\sqrt{s} = 130$ GeV per nucleon 
\cite{Phob1} as a function of $N_{part}$ and its description by the 
QGSM with percolation effects and $n_{max} = 1.67$ (solid curve) and DSM
with string fusion (dashed curve).}
\end{figure}

These data are in good agreement (Fig. 7) with our calculations in both 
QGSM and DSM including $pp$ point. Our curves are very close to a straight 
line, again due to Eq. (2) for QGSM case.

In \cite{Phob1} it was observed also the difference in 
$\eta$-distributions for the central and peripheral $Au-Au$ collisions.
Qualitatively the same difference for rapidity distributions of charged 
secondaries produced in $pp$ and $Au-Au$ collisions at $\sqrt{s} = 130$ 
GeV per nucleon is presented in Fig. 8. Some shoulder in the dashed curve 
is connected with secondary proton contribution. It is an artifact of our 
approach, because we assume that all secondaries have the transverse 
momenta equal to their averaged values.

\begin{figure}
\centerline{\epsfig{file=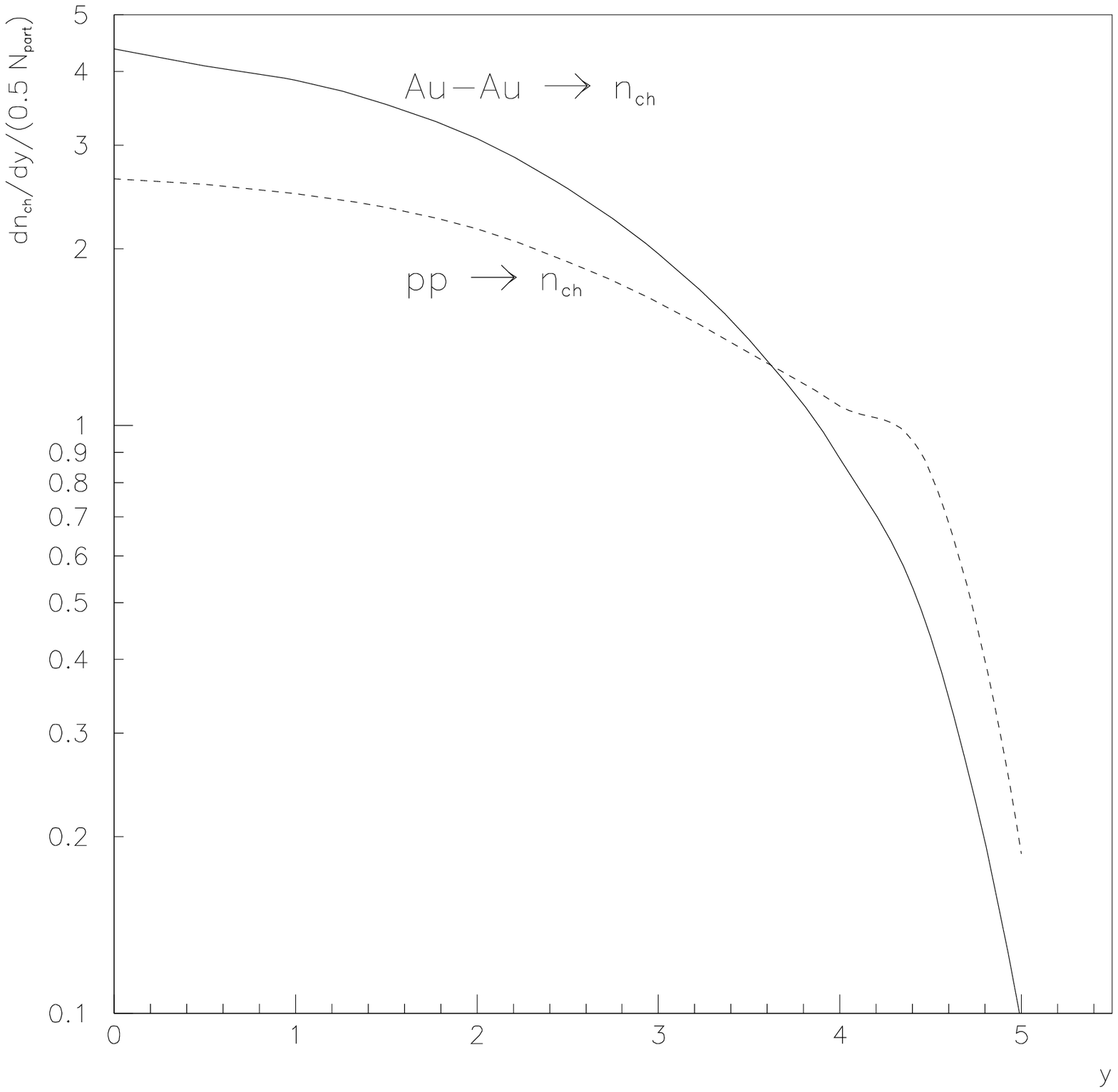,width=4in}}
\caption{Predicted by QGSM relative rapidity distributions of charged 
secondaries produced in $Au-Au$ (solid curve) and $pp$ (dashed curve) 
collisions at $\sqrt{s} = 130$ GeV per nucleon.}
\end{figure}

The observed differences for rapidity distributions and their numerical 
values are rather important. They should be connected \cite{Sh4} with 
the possibility of multiple interactions of every nucleon in heavy ion 
collision.

%\newpage

\vskip 0.9 truecm
\noindent{\bf 4. Conclusions}
\vskip 0.5 truecm

Immediately after the first data of RHIC and their comparison with CERN 
SpS data we see for the first time in high energy physics 
the pomeron (secondary particle) density saturation (percolation
effects or contribution from pomeron interactions).

With accounting for this new effects with the help of simplest estimations
and having only one free parameter ($n_{max}$ -- the maximal number of 
pomerons emitted by one nucleon) we can reproduce many features, observed 
experimentally in heavy ion collisions. We predict the independence of 
relative inclusive density $dn_{ch}/dy/(0.5 N_{part})$ in the interval 
between minimum bias and central collisions, i.e. for $A/2 < N_{part} < 
2A$ \cite{PSh}, as it is shown in Fig. 5 for mid-rapidity events. We 
describe the same behaviour ($n_{ch} \sim N_{part}$, Fig. 6) for averaged 
multiplicities practically in all rapidity region. We reproduce the 
qualitative difference between the central and peripheral (or $pp$) heavy 
ion collisions.  

However the discussed new data show some problems for our approach. 
It seems the most important are rather small ratios of antibaryon to 
baryon in mid-rapidity region. For example, the ratio of $\bar{p}/p$
yield in the central region is about 0.6 whereas our approach predicts
about 0.85 in agreement with calculations of \cite{APS}. Possibly, the 
final state interactions \cite{CS} can change the situation. In the 
opposite case it means that we overestimate the contribution of sea quarks
in the central region.

We are grateful to N.Armesto for discussions.
This paper was supported by grant NATO PSTCLG 977275.

\vskip 0.9 truecm \noindent{\bf APPENDIX. Quark and diquark distributions 
and their fragmentation functions in QGSM} 

\vskip 0.5 truecm

In the present calculations we use quark and diquark distributions in the 
proton given by the correspondent Regge behaviour \cite{KTM,Kai}. In the 
case of $n$-pomeron exchange the distributions of valence quarks and 
diquarks are softened than in the case of one-pomeron exchange due to the 
appearance of sea quark contribution. Also it is necessary to account that 
$d$-quark distribution is more soft in comparison with $u$-quark one. 
There is some freedom \cite{KTMSh} how to account for these effects. We 
use the simplest way and write diquark and quark distributions (for 
$\alpha_R = 0.5$ and $\alpha_B = -0.5$) as 
\begin{eqnarray}
u_{uu}(x,n) &=& C_{uu}x^{\alpha_R-2\alpha_B+1
}(1-x)^{\frac43(n-1)-\alpha_R}\;, \\ 
u_{ud}(x,n) &=& C_{ud}x^{\alpha_R-2\alpha_B }(1-x)^{n-1-\alpha_R}\;, \\
\\ u_{u}(x,n) &=& C_{u}x^{-\alpha_R}(1-x)^{n+\alpha_R-2\alpha_B+1}\;, \\
\\ u_{d}(x,n) &=& C_{d}x^{-\alpha_R}
(1-x)^{\frac43(n-1)+\alpha_R-2\alpha_B+1}\;, \\
\\u_{\overline{u}}(x,n) &=& u_{\overline{d}}(x,n)= C_{\overline{u}}
x^{-\alpha_R} \nonumber
\\&\times &[(1-x)^{n+\alpha_R -2\alpha_B-1} -
\delta /2(1-x)^{n+2\alpha_R-2\alpha_B-1}] \; , \; n>1\ \ ,\\
u_{s}(x,n) &=& C_{s}x^{-\alpha_R}(1-x)^{n+2\alpha_R-2\alpha_B-1} \; , \;
n>1\ \ ,
\end{eqnarray}
where $\delta =0.4$ is the relative probability to find a strange quark in
the sea. The factors $C_{i}$ are determined from the normalization 
condition 
\begin{equation}
\int_{0}^{1} u_{i}(x,n)dx = 1
\end{equation}
with sum rule
\begin{equation}
\int_{0}^{1}\sum_i u_{i}(x,n)xdx = 1\ \ .
\end{equation}

The fragmentation functions of quarks and diquarks into pions and kaons 
were changed a little in compari\-son with Refs. \cite{KTM,Sh3} to obtain 
a better agreement with the existing experimental data. We use the quark 
fragmentation functions in the form
\begin{equation}
G_{u}^{\pi^+} = G_{\bar{d}}^{\pi^+}  = G_{d}^{\pi^-} =  
G_{\bar{u}}^{\pi^-} = a_0(1-z)^{\lambda - \alpha_R}(1-0.7z) \;,
\end{equation}
\begin{equation}
G_{u}^{\pi^-} = G_{d}^{\pi^+} =  G_{\bar{d}}^{\pi^-} =  
G_{\bar{u}}^{\pi^+} = (1-z) G_{u}^{\pi^+} \;,
\end{equation}
\begin{equation}
G_{u}^{K^+} = G_{\bar{u}}^{K^-} = a_K(1 - z)^{\lambda} \;,
\end{equation}
\begin{equation}
G_u^{K^-} = G_d^{K^-} = G_d^{K^+} = G_{\bar{u}}^{K^+} = G_{\bar{d}}^{K^-}
= G_{\bar{d}}^{K^+} = G_{u}^{K^+}(1 - z) (1 - 0.95z) \;, 
\end{equation}
\begin{equation}
G_{s}^{\pi^-} = G_{s}^{\pi^+} = G_{\bar{s}}^{\pi^-} =
G_{\bar{s}}^{\pi^+} = (1-z)^{\Delta \alpha} G_{u}^{\pi^-} \;,
\end{equation}
\begin{equation}
G_{s}^{K^-} = G_{\bar{s}}^{K^+} =
a_0 (1-z)^{\lambda - \alpha_R}(1 - 0.7z) \;,
\end{equation}
\begin{equation}
G_{s}^{K^+} = G_{\bar{s}}^{K^-} = (1-z) G_{s}^{K^-} \;,
\end{equation}
with
\begin{equation}
\Delta \alpha = \alpha_{\rho} - \alpha_{\phi} = 1/2 \; ,~~~               
\lambda=2\alpha^{\prime} < p_{t}^2>=0.5 \;,~~~ a_0 = 0.73 \;,~~~a_K = 0.24  
\;.
\end{equation}

Diquark fragmentation functions have the form : 
\begin{equation}
G_{uu}^{\pi^+} = a_0(1-z)^{\lambda + \alpha_R -2\alpha_B}(1 + 2z) \; ,
G_{uu}^{\pi^-} = a_0(1-z)^{\lambda + \alpha_R -2\alpha_B + 1}(1 - 0.9z) \; 
\end{equation}
\begin{equation}
G_{ud}^{\pi^+} = G_{ud}^{\pi^-} = a_0 (1-z)^{\lambda + \alpha_R 
-2\alpha_B} (1-0.9z) \;\; ,
\end{equation}
\begin{equation}
G_{uu}^{K^+} = a_K(1-z)^{\lambda + 2(\alpha_R - \alpha_B)} \; ,
G_{ud}^{K^+} = G_{uu}^{K^+} \frac{1 + (1-z)^2}2 \;,
\end{equation}  
\begin{equation}
G_{uu}^{K^-} = G_{ud}^{K^-} = a_K(1-z)^{\lambda + 2(\alpha_R - \alpha_B) + 
1}(1 - 0.95z) 
\;.    
\end{equation}  

The probability for a process to have $n$ cut pomerons was calculated
using the quasi\--ei\-ko\-nal approximation \cite{KTM,TM}:
\begin{equation}
w_{n} = \sigma_{n}/\sum_{n=1}^{\infty}\sigma_{n}  \;  , \;
\sigma_{n} = \frac{\sigma_{P}}{nz} (1-e^{-z}
\sum_{k=0}^{n-1}\frac{z^{k}}{k!
})\ \  ,
\end{equation}
\begin{equation}
z = \frac{2C\gamma}{R^{2}+\alpha^{\prime}\xi}e^{\Delta\xi} \; ,
\; \sigma_{P} = 8\pi
\gamma e^{\Delta\xi} \; , \; \xi = \ln(s/1\ {\rm GeV}^{2})  \ \ ,
\end{equation}
with parameters
\vskip 0.3 truecm
\begin{center}
$\Delta = 0.09 \; , \; \alpha^{\prime} = 0.21\ {\rm GeV}^{-2} \; ,
\; \gamma_{pp} = 2.2\ {\rm GeV}^{-2} \; , $\newline
$R_{pp}^{2} = 3.18\ {\rm GeV}^{-2} \; , \; C_{pp} = 1.35 \;.$
\end{center}

\end{document}